\documentclass[twocolumn,showpacs,aps,prl,floatfix,superscriptaddress]{revtex4}

\usepackage{graphicx}
\usepackage{dcolumn}
\usepackage{amsmath}
\usepackage{epsfig}

\RequirePackage{xspace}

\usepackage{relsize}
\def\babar{\mbox{\slshape B\kern-0.1em{\smaller A}\kern-0.1em
    B\kern-0.1em{\smaller A\kern-0.2em R}}}

\def\epem       {\ensuremath{e^+e^-}\xspace}

\def\mumu       {\ensuremath{\mu^+\mu^-}\xspace}

\def\ccbar {\ensuremath{c\overline c}\xspace}

\def\Kbar  {\kern 0.2em\overline{\kern -0.2em K}{}\xspace}

\def\Kz    {\ensuremath{K^0}\xspace}
\def\Kzb   {\ensuremath{\Kbar^0}\xspace}
\def\KzKzb {\ensuremath{\Kz \kern -0.16em \Kzb}\xspace}
\def\Kp    {\ensuremath{K^+}\xspace}
\def\Km    {\ensuremath{K^-}\xspace}

\def\KpKm  {\ensuremath{\Kp \kern -0.16em \Km}\xspace}

\def\Dbar    {\kern 0.2em\overline{\kern -0.2em D}{}\xspace}

\def\Dz      {\ensuremath{D^0}\xspace}
\def\Dzb     {\ensuremath{\Dbar^0}\xspace}
\def\DzDzb   {\ensuremath{\Dz {\kern -0.16em \Dzb}}\xspace}
\def\Dp      {\ensuremath{D^+}\xspace}
\def\Dm      {\ensuremath{D^-}\xspace}

\def\DpDm    {\ensuremath{\Dp {\kern -0.16em \Dm}}\xspace}

\def\Bbar    {\kern 0.18em\overline{\kern -0.18em B}{}\xspace}

\def\BB      {\ensuremath{B\Bbar}\xspace} 
\def\Bz      {\ensuremath{B^0}\xspace}
\def\Bzb     {\ensuremath{\Bbar^0}\xspace}
\def\BzBzb   {\ensuremath{\Bz {\kern -0.16em \Bzb}}\xspace}
\def\Bu      {\ensuremath{B^+}\xspace}
\def\Bub     {\ensuremath{B^-}\xspace}
\def\Bp      {\ensuremath{\Bu}\xspace}

\def\BpBm    {\ensuremath{\Bu {\kern -0.16em \Bub}}\xspace}

\def\jpsi     {\ensuremath{{J\mskip -3mu/\mskip -2mu\psi\mskip 2mu}}\xspace}

\mathchardef\Upsilon="7107
\def\Y#1S{\ensuremath{\Upsilon{(#1S)}}\xspace}% no space before {...}!

\def\FourS {\Y4S}

\mathchardef\Deltares="7101
\mathchardef\Xi="7104
\mathchardef\Lambda="7103
\mathchardef\Sigma="7106
\mathchardef\Omega="710A

\def\Deltabar{\kern 0.25em\overline{\kern -0.25em \Deltares}{}\xspace}
\def\Lbar{\kern 0.2em\overline{\kern -0.2em\Lambda\kern 0.05em}\kern-0.05em{}\xspace}
\def\Sigbar{\kern 0.2em\overline{\kern -0.2em \Sigma}{}\xspace}
\def\Xibar{\kern 0.2em\overline{\kern -0.2em \Xi}{}\xspace}
\def\Obar{\kern 0.2em\overline{\kern -0.2em \Omega}{}\xspace}
\def\Nbar{\kern 0.2em\overline{\kern -0.2em N}{}\xspace}
\def\Xb{\kern 0.2em\overline{\kern -0.2em X}{}\xspace}

\def\BR         {{\ensuremath{\cal B}\xspace}}

\def\mes        {\mbox{$m_{\rm ES}$}\xspace}

\def\DeltaE     {\mbox{$\Delta E$}\xspace}

\newcommand{\tev}{\ensuremath{\mathrm{\,Te\kern -0.1em V}}\xspace}
\newcommand{\gev}{\ensuremath{\mathrm{\,Ge\kern -0.1em V}}\xspace}
\newcommand{\mev}{\ensuremath{\mathrm{\,Me\kern -0.1em V}}\xspace}
\newcommand{\kev}{\ensuremath{\mathrm{\,ke\kern -0.1em V}}\xspace}
\newcommand{\ev}{\ensuremath{\mathrm{\,e\kern -0.1em V}}\xspace}
\newcommand{\gevc}{\ensuremath{{\mathrm{\,Ge\kern -0.1em V\!/}c}}\xspace}
\newcommand{\mevc}{\ensuremath{{\mathrm{\,Me\kern -0.1em V\!/}c}}\xspace}
\newcommand{\gevcc}{\ensuremath{{\mathrm{\,Ge\kern -0.1em V\!/}c^2}}\xspace}
\newcommand{\mevcc}{\ensuremath{{\mathrm{\,Me\kern -0.1em V\!/}c^2}}\xspace}
\def\cm   {\ensuremath{\rm \,cm}\xspace}

\def\mm   {\ensuremath{\rm \,mm}\xspace}

\def\mum  {\ensuremath{\,\mu\rm m}\xspace}%% mu meter 
\def\invfb   {\ensuremath{\mbox{\,fb}^{-1}}\xspace}

\def\mus  {\ensuremath{\rm \,\mus}\xspace}

\def\mus        {\ensuremath{\,\mu{\rm s}}\xspace}    %% microsecond
\def\rad{\ensuremath{\rm \,rad}\xspace}

\def\to                 {\ensuremath{\rightarrow}\xspace}

\def\pep2{PEP-II}

\newcommand{\chisq}{\ensuremath{\chi^2}\xspace}

\def\gsim{{~\raise.15em\hbox{$>$}\kern-.85em
          \lower.35em\hbox{$\sim$}~}\xspace}
\def\lsim{{~\raise.15em\hbox{$<$}\kern-.85em
          \lower.35em\hbox{$\sim$}~}\xspace}

\xspace

\newcommand{\jprlBase}       {Phys.\ Rev.\ Lett.\xspace}
\newcommand{\jprBase}        {Phys.\ Rev.\xspace}
\newcommand{\jplBase}        {Phys.\ Lett.\xspace}

\newcommand{\nimBaseC}       {Nucl.\ Instr.\ and Methods\xspace}

\newcommand{\nim}       [1]  {\nimBaseC~{\bf #1}}
\newcommand{\plb}       [1]  {\jplBase\ B~{\bf #1}}

\newcommand{\jprl}      [1]  {\jprlBase\ {\bf #1}}
\newcommand{\jprd}      [1]  {\jprBase\ D~{\bf #1}}
\def\jetset74   {\mbox{\tt Jetset \hspace{-0.5em}7.\hspace{-0.2em}4}\xspace}

\def\pstar {\ensuremath{{p^*}}}
\def\lambar {\ensuremath{{\overline{\Lambda}}}}

\def\JpL {\ensuremath{{B^+\to\jpsi p \overline{\Lambda}}}}
\def\Cback {\ensuremath{{0.21\pm0.14}}}
\def\Cprob {\ensuremath{{2.5\times10^{-4}}}}
\def\Cfcor {\ensuremath{{0.0054\pm0.0035}}}
\def\Cbf {\ensuremath{{(12 ^{+9}_{-6})}}}
\def\Cbffull {\ensuremath{{(11.6 ^{+8.5}_{-5.6})}}}
\def\Cbfstat {\ensuremath{{(11.6 ^{+7.4}_{-5.3})}}}
\def\Cul {\ensuremath{{26}}}
\def\Ceff {\ensuremath{{0.049\pm0.009}}}

\def\Jpp {\ensuremath{{B^0\to\jpsi p \overline{p}}}}
\def\Nback {\ensuremath{{0.64\pm0.17}}}
\def\Nfcor {\ensuremath{{0.0051\pm0.0013}}}
\def\Neff {\ensuremath{{0.184\pm0.024}}}
\def\Nul {\ensuremath{{1.9}}}

\newcommand{\BABARPubYear}    {03}
\newcommand{\BABARPubNumber}  {007}

\newcommand{\SLACPubNumber} {9690}

\begin{document}

\title{\begin{flushleft}
\babar-PUB-\BABARPubYear/\BABARPubNumber \\
SLAC-PUB-\SLACPubNumber \\
\end{flushleft}
Evidence for \JpL\ and Search for \Jpp}

%% author list as of 01-Feb-2003 (555 authors)
%
\author{B.~Aubert}
\author{R.~Barate}
\author{D.~Boutigny}
\author{J.-M.~Gaillard}
\author{A.~Hicheur}
\author{Y.~Karyotakis}
\author{J.~P.~Lees}
\author{P.~Robbe}
\author{V.~Tisserand}
\author{A.~Zghiche}
\affiliation{Laboratoire de Physique des Particules, F-74941 Annecy-le-Vieux, France }
% ...
\collaboration{The \babar\ Collaboration}
\noaffiliation

\date{\today}

\begin{abstract}
We have performed a search for the decays \JpL\ and \Jpp\ in a data
set of $(88.9 \pm 1.0)\times 10^6$ \FourS\ decays collected by the
\babar\ experiment at the PEP-II 
\epem\ storage ring at the Stanford
Linear Accelerator Center.  
Four charged $B$
candidates have been observed 
with an expected background of \Cback\ events. The
corresponding
branching fraction is $\Cbf\times 10^{-6}$, where statistical 
and systematic uncertainties have been combined. 
The result can be
interpreted as a 90\% confidence level 
(CL) upper limit of $\Cul \times 10^{-6}$. 
We also find one \Bz\ candidate, with an expected background of
\Nback\ events, implying a  
90\% CL upper limit of $\Nul\times 10^{-6}$. 

\end{abstract}


\maketitle

\begin{figure}
\centering
\includegraphics[width=3in]{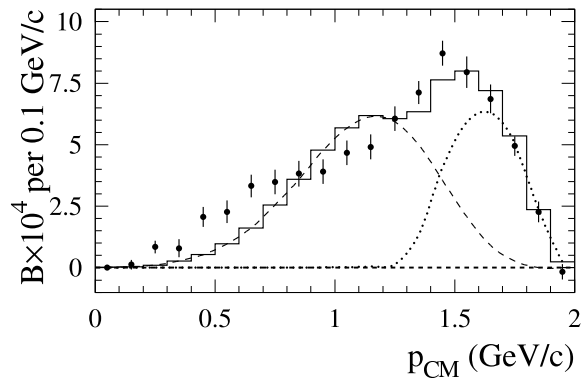}
\caption{Center-of-mass momentum 
of \jpsi\ mesons produced directly in
$B$ decays 
(points). The histogram is the sum of
the color-octet component from a recent NRQCD 
calculation \cite{ref:benekep} (dashed
line), which includes multi-body final states,
and the color-singlet $\jpsi K^{(*)}$ component from 
simulation \cite{ref:jetset} (dotted line).} 
\label{fig:directjpsi}
\end{figure}

Studies of the inclusive production of charmonium mesons in $B$ decays
at the \FourS\ resonance 
have been published by CLEO \cite{ref:cleo}
and \babar\ \cite{ref:incjpsi}, and preliminary results have
been presented by Belle \cite{ref:schrenk}. 
One of the interesting features observed by
all three collaborations is an excess of \jpsi\ mesons at low
momentum in the \epem\ center-of-mass frame, $p_{CM}$, when 
compared to distributions predicted by non-relativistic
QCD calculations \cite{ref:benekep}. 
Figure~\ref{fig:directjpsi} (from Ref.~\cite{ref:incjpsi})
shows $p_{CM}$ for \jpsi\ mesons produced in $B$ decay after
subtraction of the component due to the decay of heavier charmonium
states. The excess below 0.8\gevc\ corresponds to a branching fraction
of approximately 
$6 \times 10^{-4}$, 8\% of the total direct \jpsi\ production. 

Possible sources of the excess 
include an intrinsic charm component of the $B$ \cite{ref:hou}  
or the production of 
an $s{\overline d}g$ hybrid \cite{ref:eilam} in conjunction with a
\jpsi. 
Another possibility \cite{ref:brodsky}
is that the excess is from decays of the form $B \to
\jpsi$ {\em baryon anti-baryon}. 
The rate of these decays could
be enhanced by the intermediate production of an exotic state allowed by
QCD but not yet observed, 
including nuclear-bound quarkonium (a \ccbar\ pair bound to a
nucleon), baryonium (a baryon-antibaryon bound state), or 
a pentaquark (a baryon containing five quarks). 
If such resonances were narrow, the other particle in the decay would be
monoenergetic in the $B$ rest frame. Note that the \jpsi\ spectrum in
Fig.~\ref{fig:directjpsi} would not directly display such narrow
distributions because it 
is measured in the \epem\ center-of-mass frame. The
difference between $p_{CM}$ and \pstar, the \jpsi\ momentum in the $B$
rest frame, has an RMS of 0.12\gevc\ due to the motion of the $B$.

This Letter presents searches for the decays \JpL\ and
\Jpp\ in a sample of 81.9\invfb\ collected by 
the \babar\ detector. Note that the latter decay is Cabibbo suppressed
relative to the former. 
Charge conjugation is implied throughout.

\babar\ operates at the 
PEP-II \epem\ storage ring, which collides 9.0\gev\ electrons on 
3.1\gev\ positrons to 
create a 
center-of-mass system with energy 10.58\gev\ moving along the $z$
axis with a Lorentz boost of $\beta\gamma = 0.55$. 
\FourS\ production makes up 
approximately 23\% of the total hadronic cross
section.  

The \babar\ detector is described in detail in
Ref.~\cite{ref:babarnim}.  
The trajectories of charged particles are reconstructed and their 
momenta measured with two detector systems located in a
1.5-T solenoidal magnetic field: a five-layer, double-sided silicon
vertex tracker (SVT) and a 40-layer drift chamber (DCH).   
The tracking fiducial
volume covers the polar angular region $0.41 < \theta < 2.54$\rad, 
which is 86\% of the solid angle in the center-of-mass frame.
The transverse momentum resolution
is 0.49\% at 0.3\gevc\ and 0.59\% at 1\gevc.

The energies deposited by charged tracks and photons are measured by a
CsI(Tl) calorimeter (EMC) in the fiducial volume
$0.41 < \theta < 2.41$\,rad (84\% of the center-of-mass solid angle) 
with energy resolution at 1\gev\ of 2.6\%.
Muons are detected in
the IFR, a multilayer device of 
resistive 
plate chambers located in the flux return of the solenoid.   The DIRC, 
a Cherenkov radiation detector, is used to identify 
charged particles.  

We select $B$ candidates of interest in a \BB-enriched sample. Events
in the sample are required to have visible energy $E$ greater than
4.5\gev\ and a ratio of the second to the 
zeroth Fox-Wolfram moment \cite{ref:fox}, $R_2$, 
less than 0.5.  Both $E$ and $R_2$ are
calculated from  
tracks and neutral energy deposits in the respective fiducial volumes
noted 
above.  The same tracks are used to construct a primary event vertex,
which is required to be located within 6\cm\ of the beam spot in $z$
and within 0.5\cm\ of the beam line. The beam spot RMS size is
approximately 0.9\cm\ in $z$, 120\mum\ horizontally, and 
5.6\mum\ vertically.

There must be at least three tracks in the fiducial volume satisfying
the following quality criteria: they must have transverse momentum
greater than 0.1\gevc, momentum less than 10\gevc, at least 12 hits in
the DCH, and approach within 10\cm\ of the 
beam spot in $z$ and within
1.5\cm\ of the beam line.

Studies with simulated data indicate that 
these criteria are satisfied by 96\% 
of generic \BB\ events. 

\JpL\ candidates are formed by combining \jpsi, proton, and 
\lambar\ candidates.
\jpsi\ candidates must have mass in the range 2.950--3.130\gevcc\ if
reconstructed in the \epem\ final state or 3.060--3.130\gevcc\ in
\mumu. 

One of the two electrons from the \jpsi\ must satisfy the following
(``tight'') requirements. It must have an energy deposit in the EMC 
between 89\% and
120\% of its momentum, a Cherenkov angle 
in the DIRC within $3\sigma$ of expectation for an electron,
a lateral moment of the energy 
deposit \cite{ref:lat}, LAT, between 0.1 and 0.6, an
$A_{42}$ Zernike moment \cite{ref:zern} less than 0.11, and an 
energy loss in the DCH consistent with expectation.
Less stringent (``loose'') requirements are imposed in the selection
of the second electron: we require
an energy deposit in the EMC of at
least 65\% of its momentum and place a
less restrictive requirement on DCH energy, with no requirements 
on LAT or $A_{42}$.
Whenever possible,
photons radiated by an electron traversing material prior to the DCH
(0.04 radiation lengths at normal incidence) 
are combined with the track \cite{ref:incjpsi}. 

At 1.5\gevc, a typical lepton momentum,
the tighter criteria have an efficiency of 91\% with a pion
misidentification probability of 0.13\%. The looser criteria give 98\%
efficiency with 3\% pion misidentification. 

Muon candidates must deposit less than 0.5\gev\ in the EMC
(2.3 times the minimum-ionizing peak) and have a pattern of
hits in the IFR consistent with the trajectory of a muon.
The total amount of material penetrated must be 
greater than 2 interaction
lengths and must be within 2 interaction lengths of the value
expected for a muon. The 
muon identification efficiency at 1.5\gevc\ is 77\% with a pion
misidentification probability of 11\%.

Proton candidates are selected with a likelihood method that uses the
energy deposited in the SVT and the DCH, and the  
Cherenkov angle and number of photons observed in the
DIRC. They are also required to fail the tight electron identification
criteria. At a typical momentum of 300\mevc, 
the selection efficiency is 
greater than 98\%
with a kaon misidentification probability less than 1\%.

The \lambar\ is reconstructed from a proton, which must satisfy the
above criteria, and an oppositely charged track, assumed to be a
pion. It 
must have mass between 1.10 and 1.14\gevcc, and a vertex that
is separated from the \jpsi\ vertex by at least 2\mm. 
The angle between the \lambar\ momentum and the vector from the
\jpsi\ vertex to the \lambar\ vertex must be less than $90^\circ$
in the laboratory frame.

Geometrical vertex fits are performed on the resulting 
\Bp\ candidates, of which  
approximately 68\% are rejected by a requirement on the quality of
the fit. 

\Jpp\ candidates are formed from \jpsi\ candidates and an 
oppositely-charged pair of proton candidates. Approximately 
83\% of resulting candidates fail a requirement on the quality of a
vertex fit. 

We use two nearly-independent
kinematic variables \cite{ref:babarnim} 
to categorize $B$ candidates: the difference between the 
reconstructed and expected energy of the $B$ candidate in the 
\epem center-of-mass frame,  
$\DeltaE  =  (q_\Upsilon \cdot q_B - s/2)/\sqrt{s}$, 
and the beam-energy substituted mass, 
$\mes  =  \sqrt{ (0.5s + \vec{p}_B \cdot
\vec{p}_\Upsilon)^2/E^2_\Upsilon - p^2_B}$. 
The four-momentum of the \epem\ initial state, obtained from the beam
momenta, is $q_\Upsilon = (E_\Upsilon,\vec{p}_\Upsilon)$, 
and $s \equiv |q_\Upsilon|^2$.
The four-momentum of the reconstructed $B$
candidate, $q_B = (E_B,\vec{p}_B)$, is
found by summing the four-momenta of the
three daughters, with daughter masses constrained to 
accepted values \cite{ref:pdg2002}.

The ``analysis window'' AW is defined by 
$5.2 < \mes < 5.3$\gevcc\  and 
$-0.10 < \DeltaE < 0.25$\gev\ (\Bp\ candidates) and
$-0.25 < \DeltaE < 0.25$\gev\ (\Bz\ candidates).
The \DeltaE\ range is smaller for the charged candidates due to a
kinematic cutoff in the \JpL\ decay. 
Only candidates in the AW are considered in the analysis. 
Approximately 15\% of \Bp\ events 
and 1.5\% of \Bz\ events contain more than one candidate, in which 
case we select the one with the lowest $|\DeltaE|$.

For signal events, $\langle \DeltaE \rangle \approx 0$ 
and $\langle \mes \rangle \approx M_B$. 
We define a signal ellipse by $\left[ \left( \mes - M_B \right) /
\sigma_m \right]^2 +  
\left[ \DeltaE / \sigma_E \right]^2 < S^2$, where the resolutions 
$\sigma_m$ and $\sigma_E$ are estimated from simulated data
to be 3.1\mevcc\ and 6.5\mev, respectively, for \JpL, and 
2.7\mevcc\ and 5.5\mev\ for \Jpp.
$S = 2.4$ for \JpL\ and $S=2.2$ for \Jpp. 

The selection criteria for charged and neutral $B$ candidates,
including the values for $S$, have been chosen to minimize the 90\%
CL upper limit expected in the absence of real signal, based on
simulated signal and background events. 
Approximately 90\% of
the background events satisfying the criteria 
are combinatorial \BB,
in which tracks from the decays of both $B$ mesons are used to form
the candidate. The rest are continuum (non-\BB) events. 
Both components are distributed throughout the AW, 
and neither peaks in the the signal of either \DeltaE\ or \mes. 

We use simulated \JpL\ and \Jpp\ events
to measure the selection efficiency. The simulation does not include
exotic QCD bound states. 
We study the accuracy of the simulation of the detector response by 
comparing data and simulated background events in 
samples similar to the final selection. We compare the number of
\jpsi\ mesons reconstructed in \Jpp\ candidates in which only one proton 
satisfies the identification criteria, and we compare
the number of \lambar\ baryons reconstructed in \JpL\ candidates in
which the 
proton daughter of the \Bp\ is required to fail the criteria. Based on
these studies, we apply multiplicative corrections to the efficiency
of $0.97\pm0.06$ for \jpsi\ reconstruction and
$0.86\pm0.14$ for \lambar\ reconstruction. 
We also compare the distributions of the
\chisq\ of the $B$ vertex for candidates satisfying all other criteria
and obtain corrections of $0.98\pm0.02$ for \JpL\ and
$0.90\pm 0.10$ for \Jpp.

The efficiency for \JpL, with the \jpsi\ decaying to \epem\ or
\mumu\ and \lambar\ decaying to $\overline{p} \pi^+$, 
is \Ceff. The 18\% fractional uncertainty includes 16\% from
\lambar\ reconstruction, 6\% from the \jpsi,  3\% from statistical
uncertainty in the simulation, 2\% from the 
\chisq\ correction, and 1\% uncertainty on
proton reconstruction efficiency. Approximately 25\% of signal events
satisfying all other criteria are reconstructed outside the signal
ellipse. 

The efficiency for \Jpp\ with the \jpsi\ decaying to \epem\ or
\mumu\ is \Neff. The 13\% uncertainty includes 6\%
from \jpsi\ reconstruction,  2\% for 
statistical uncertainty in the simulation, 11\% for
the \chisq\ correction, and 3\% 
for proton reconstruction.

\begin{figure}
\includegraphics[width=\linewidth]{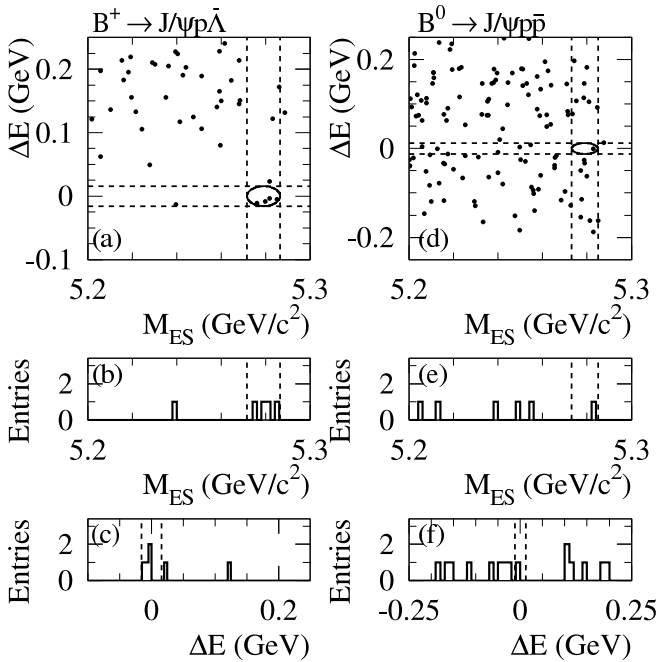}
\caption{(a) Distribution of \JpL\ candidates in the 
\DeltaE-\mes\ plane, with the signal ellipse and its projection in
each dimension 
(dashed lines). Histogram of candidates within marked bands  in (b)
\mes\ and (c) \DeltaE. Plots (d)--(f) show similar quantities for
\Jpp. 
}
\label{fig:ProjData}
\end{figure}

We use
world average values \cite{ref:pdg2002}  
for $\BR(\jpsi\to\epem)$,  $\BR(\jpsi\to\mumu)$, and 
$\BR(\Lambda \to p\pi^-)$.

We estimate the mean expected background in the signal ellipse
($\mu_B$) 
from the number $N_A$ elsewhere in the AW: $\mu_B = f\cdot N_A$.
We obtain $f$, the proportionality constant, from a larger 
sample in which only one proton satisfies the proton
identification criteria. 
We perform a Kolmogorov test \cite{ref:kolmo} to verify that 
the distribution of candidates in the
\DeltaE-\mes\ plane is similar to the standard selection. 
Comparing the regions outside the ellipse, the test 
gives a probability of 0.52 for \JpL\ and 0.36 for
\Jpp. We obtain $f = \Cfcor$ (\Bp) and $f = \Nfcor$ (\Bz). The
uncertainties are largely statistical, but include a component 
(16\% for \Bp\ and 2\% for \Bz) due to differences in the number of
events with multiple candidates. 

For \JpL, $N_A = 39$, implying an expected background 
of \Cback\ events. We observe four candidates in the 
signal ellipse (Fig.~\ref{fig:ProjData}). 
The probability of observing $\ge 4$ candidates
when expecting \Cback\ is \Cprob. Three of the four are positively
charged. Two of the four \jpsi\ mesons decay to \epem\ and two to
\mumu. 

To interpret this result as a $B^+$ branching fraction \BR,
we
undertake a Bayesian analysis with a uniform prior above zero. We
define the likelihood for \BR\ as the probability of observing exactly
four events, including uncertainties on the expected
background, signal efficiency, secondary branching fractions, and
number of \FourS\ decays, $(88.9 \pm 1.0)\times 10^6$.
We assume the branching fractions 
$\BR(\FourS\to \BpBm) = \BR(\FourS\to\BzBzb) = 0.5$. 

The central value for \BR\ is the peak of the likelihood
function. 
We obtain ``$\pm 1 \sigma$'' uncertainties from a 
confidence interval that encloses 68.3\% of the area of the
likelihood function, selected such that 
the likelihoods for all values of
\BR\ in the interval are
larger than the likelihoods outside. The result is
$\BR(\JpL) = \Cbffull \times 10^{-6}$. We similarly obtain a  
90\% CL upper limit of $\Cul\times 10^{-6}$. 

If we consider only the 
statistical uncertainty, the result would be 
$\BR(\JpL) = \Cbfstat \times 10^{-6}$. Subtracting these uncertainties
in quadrature would indicate contributions from systematic errors of 
$4.2\times 10^{-6}$ and $1.8\times 10^{-6}$ on the upper and lower
sides respectively. The systematic error arises almost entirely from
the uncertainty on the signal efficiency. 

\begin{figure}
\includegraphics[width=\linewidth]{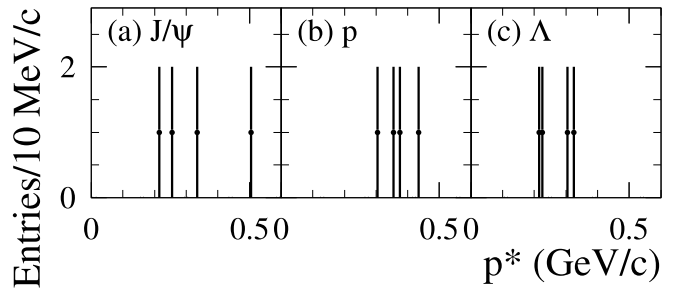}
\caption{Momentum in the $B^+$ rest frame of the (a) \jpsi, (b)
proton, and (c) \lambar\ daughters of the four \JpL\ candidates.
}
\label{fig:pstarJpL}
\end{figure}

The creation of a narrow QCD exotic bound state 
as an intermediate resonance in the
\Bp\ decay would be reflected as a narrow \pstar\ distribution
of the other decay daughter.
We do not observe any significant clustering in the 
\pstar\ distributions of the \jpsi, proton, or \lambar\ daughters of
the four \Bp\ candidates (Fig.~\ref{fig:pstarJpL}). The resolution in
\pstar\ is $\sigma \sim 20$\mevc. 

For \Jpp, there are 126 events outside the signal ellipse, indicating
an expected background of \Nback\ events, and one event in the ellipse.
Following the procedure described for \JpL, and again assuming a
uniform prior above 0, we obtain 
$\BR(\Jpp) < \Nul \times 10^{-6}$ (90\% CL). This limit is dominated by
statistical uncertainty. 

In summary, 
we observe four \JpL\ candidates in a data set of 
$(88.9 \pm 1.0)\times 10^6$ \FourS\ decays.
The probability of the expected charged $B$ background,
\Cback\ events, producing $\ge 4$ events is \Cprob. The branching
fraction is $\Cbf\times 10^{-6}$, where the uncertainty includes both
statistical and systematic components. 
This result can be interpreted as
a 90\% CL upper limit of $\Cul \times 10^{-6}$.

We observe one \Jpp\ candidate with an expected background of
\Nback, and determine a 90\% CL upper limit of 
$\Nul\times 10^{-6}$ on the branching fraction. 

Neither final state makes a significant
contribution to the 
observed excess of \jpsi\ mesons in inclusive $B$ decay.
The momentum
distributions of the \Bp\ daughters do not 
provide evidence for
QCD exotic particles produced as narrow intermediate states.

% Input the pubboard acknowledgments file
We are grateful for the excellent luminosity and machine conditions
provided by our \pep2\ colleagues, 
and for the substantial dedicated effort from
the computing organizations that support \babar.
The collaborating institutions wish to thank 
SLAC for its support and kind hospitality. 
This work is supported by
DOE
and NSF (USA),
NSERC (Canada),
IHEP (China),
CEA and
CNRS-IN2P3
(France),
BMBF and DFG
(Germany),
INFN (Italy),
FOM (The Netherlands),
NFR (Norway),
MIST (Russia), and
PPARC (United Kingdom). 
Individuals have received support from the 
A.~P.~Sloan Foundation, 
Research Corporation,
and Alexander von Humboldt Foundation.

\end{document}